\documentclass[useAMS,usenatbib]{mn2e}
\usepackage{natbib}
\usepackage{graphicx}

\title[On the structure of tidal tails]{On the structure of tidal tails}

\author[A.H.W. K\"upper, A. Macleod and  D.C. Heggie]{Andreas
  H.W. K\"upper$^{1}$\thanks{E-mail: \mbox{akuepper@astro.uni-bonn.de} (AHWK);
    \mbox{A.Macleod-6@sms.ed.ac.uk} (AM);  \mbox{d.c.heggie@ed.ac.uk} (DCH)}, Andrew
  Macleod$^2$ and Douglas C. Heggie$^{3}$\\
$^{1}$Argelander Institute for Astronomy (AIfA), Auf dem H\"ugel 71, D-53121 Bonn, Germany\\
$^2$University of Edinburgh, School of Physics, King's
Buildings, Edinburgh EH9 3JZ, UK\\
$^{3}$University of Edinburgh, School of Mathematics and Maxwell
Institute for Mathematical Sciences, King's
Buildings, Edinburgh EH9 3JZ, UK}

\begin{document}

\date{Accepted \ldots. Received \ldots; in original form \ldots}

\pagerange{\pageref{firstpage}--\pageref{lastpage}} \pubyear{2008}

\maketitle

\label{firstpage}

\maketitle

\begin{abstract}
We examine the longitudinal distribution of the stars escaping from a
cluster along tidal tails. Using both theory and simulations, we show
that, even in the case of a star cluster in a circular galactic orbit,
when the tide is steady, the distribution exhibits maxima at a
distance of many tidal radii from the cluster.
\end{abstract}


\section{Introduction}

Stars escaping from a star cluster in the gravitational field of a
galaxy form extended {\sl tidal tails}.  These have been observed in
the Milky Way \citep{grillmair1995,lehmann1997,kharchenko1997,leon2000,Odenkirchen01,rockosi2002,lee2003,belokurov2006}
and have often been modelled
\citep{combes1999,johnston1999,yim2002,Dehnen04,koch2004,lee2004,odenkirchen2003,dimatteo2005,capuzzo2005,lee2006,chumak2006,choi2007,fellhauer2007,montuori2007}. They
exhibit significant longitudinal structure, i.e. clumps, or over- and
under-densities, which are usually attributed to the influence of
gravitational shocks, e.g. encounters with spiral arms or giant
molecular clouds, or passages through a galactic disk or past a
galactic bulge, or simply motion in a triaxial potential.  Here we show that clumps also arise when the tidal
field is static, in the simplest case of a star cluster in a circular
orbit about an axisymmetric galactic potential.  In the next section
we describe a theoretical approach to the problem, which we verify in
$N$-body calculations in Section 3.  The final section is a slightly
extended summary with discussion.

\section{Theory}\label{sec:theory}
\begin{figure*}
  \centering
  \includegraphics[width=168mm]{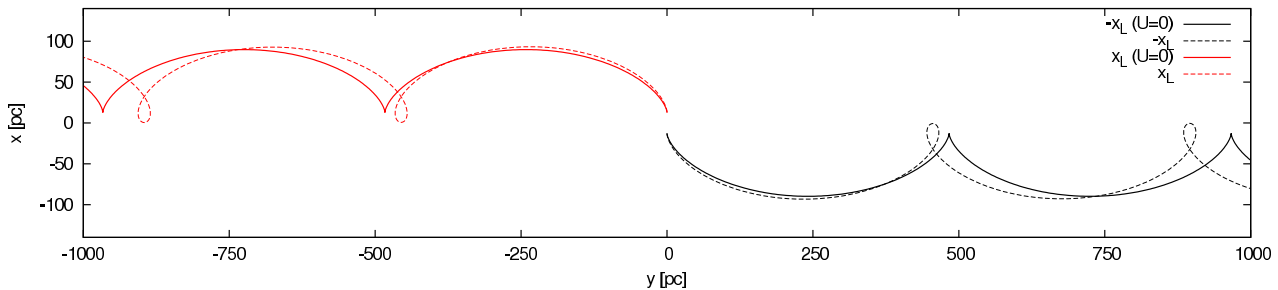}
\caption{Comparison of eqs.(\ref{eq:xapprox}),(\ref{eq:yapprox})
  (solid lines) with the solution of eqs.(\ref{eq:xeom}),(\ref{eq:yeom}) (dashed lines), with the
  point-mass approximation for $U$, with the same initial conditions
  starting at $\pm x_L$ (except for a small outward $x$-component of
  velocity).  The units and tidal parameters are such that $M = 1000\,\mbox{M}_\odot$ and $x_L = 12.9$ pc.}\label{fig:orbitplot}
\end{figure*}

As in standard theory \citep{Chandrasekhar42} we describe the position of a
star in a rotating, accelerated coordinate system with origin at the
centre of the cluster, and axes aligned towards the galactic
anticentre, in the direction of galactic motion of the cluster, and
orthogonal to the plane of motion of the cluster, respectively.  Then
the equations of motion are
\begin{eqnarray}
  \ddot x - 2\Omega \dot y + (\kappa^2 - 4\Omega^2)x &=& -U_x\label{eq:xeom}\\
  \ddot y + 2\Omega \dot x  &=& -U_y\label{eq:yeom}\\
  \ddot z + \omega_z^2 z &=& -U_z\label{eq:zeom},
\end{eqnarray}
where $\Omega$ is the angular velocity of the cluster around the
galaxy, $\kappa$ is the epicyclic frequency,  $\omega_z$ is the
frequency of motions orthogonal to the plane of motion of the cluster,
and $U$ is the gravitational potential of the cluster stars.

We are interested in escaping stars, and so we shall adopt a
point-mass approximation for $U$.  We let
 $U = \displaystyle{-GM/r}$, where $G$ is the gravitational
constant, $M$ is the total mass of the cluster and $r =
\sqrt{x^2+y^2+z^2}$.  There are equilibria at the Lagrangian points
$(\pm x_L,0,0)$, where 
$$
x_L = \displaystyle{\left(\frac{GM}{4\Omega^2-\kappa^2}\right)^{1/3}}.\label{eq:xl}$$
The total effective energy of a star (per unit mass) is 
$$
E = \frac{1}{2}v^2 + \Phi,
$$
where $v$ is the star's speed and the 
effective potential energy is 
$$
\Phi =
\displaystyle{\frac{1}{2}((\kappa^2 - 4\Omega^2)x^2 + \omega_z^2z^2) -
  \frac{GM}{r}}.
$$ 
The Lagrangian points are saddle points of $\Phi$,
where its value is $\Phi_L =
-\displaystyle{\frac{3}{2}\frac{GM}{x_L}}$.  This is also the escape
energy from the potential well of the cluster in the galactic field.

Except for high-speed escapers created in few-body encounters, stars
escape as a result of two-body encounters and then an escaper usually
has an energy only slightly above the escape energy $\Phi_L$ \citep{hayli1970}.
Therefore such escapers pass close to one of the Lagrange points at slow speed.
Thereafter the acceleration due to the cluster rapidly diminishes, and
the motion of the escaping star can be well approximated by
eqs.(\ref{eq:xeom})-(\ref{eq:zeom}) with the right sides set equal to
zero. The solution starting at time $t=0$ at position $(x_L,0,0)$
with zero velocity is then easily found to be
\begin{eqnarray}
  x &=& \frac{4\Omega^2}{\kappa^2}x_L + \left(1 -
  \frac{4\Omega^2}{\kappa^2}\right)x_L\cos\kappa t\label{eq:xapprox}\\
y &=& -\frac{2\Omega}{\kappa}\left(1 -
  \frac{4\Omega^2}{\kappa^2}\right)x_L(\sin\kappa t - \kappa t)\label{eq:yapprox}\\
z &=& 0.
\end{eqnarray}
The solution for the stars starting at $(-x_L,0,0)$ can be found as easily.

After leaving the Lagrange point the solution traces an oscillatory path along the tidal tail and comes to rest again at the cusp of the orbit, which is reached when 
\begin{equation}
t = t_C = 2\pi/\kappa.    
\end{equation}
At this time  the location of the star is $y = -y_C$
where 
$$
y_C = -\frac{4\pi\Omega}{\kappa}\left(1-\frac{4\Omega^2}{\kappa^2}\right)x_L.\label{eq:yc}
$$

Now let us suppose there is a stream of stars following this motion.
Because the $y-$component of velocity is zero at $y = -y_C$, there is
a peak (actually, an infinite peak) in the distribution of the values of $y$ at this location.  In
the evolution of a star cluster, however, it will not be visible until
the first escapers reach this location, at around time $t_C$.
Furthermore, as the star cluster loses mass by escape, $x_L$
decreases, and the location of this peak moves closer to the cluster
as the cluster dissolves.  The
existence of this peak, and the time when it begins to become
apparent, are the two main predictions of this theory.

Now we discuss the extent to which the solution we have used is a
satisfactory approximation to the motion of slow escapers from a star
cluster. There are two factors to consider.
\begin{enumerate}
\item The
effect of neglecting the field of the cluster is shown in
Fig.\ref{fig:orbitplot}.  What will be important in what follows is
the $y$-coordinate of the first cusp after the star leaves the
Lagrange point, and it is clear that this is well approximated. 
\item   Our other main approximation is the assumption that the star leaves the
Lagrange point with a very small velocity.  This cannot be tested thoroughly
without recourse to an $N$-body simulation, which we postpone to
Sec.\ref{sec:numerical}.  But we can develop a feel for the result by
changing the initial conditions, e.g. by adding a small initial
value to $y$, of order $x_L$.  Then the cusp is displaced in $y$ by the same
amount, so that the relative displacement of the cusp is
$$
\frac{x_L}{y_C} =
\frac{1}{4\pi}\frac{\kappa^3}{\Omega(4\Omega^2-\kappa^2)},
$$
which is $1/(12\pi)$ in the case of a point-mass potential in which
$\Omega = \kappa$.  While this is not large, it shows that the density
enhancement at $y\simeq-y_C$ will be smoothed out.  Similar results
are obtained if we vary other initial conditions appropriately, though
usually the cusp disappears, and may be replaced by a small epicyclic
loop.  This calculation also shows that the distance of the density
enhancement from the star cluster is of order $12\pi$ tidal radii,
again in the case of a point-mass galactic potential.
\end{enumerate}

Now we consider the number of stars between the cluster and the first
density enhancement (near $y = -y_C$).  This is occupied by stars
which escape in a time interval of order $t_C$.  If we assume that a
constant fraction, $\mu$,  of stars escapes in each half-mass
relaxation time (see, for example, \citealt{Spitzer87}), the relevant number of escaping stars is 
$$
N_{esc} = \mu N\frac{t_C}{t_{rh}},
$$
where 
$$
t_{rh} = \frac{0.138N^{1/2}r_h^{3/2}}{G^{1/2}m^{1/2}\log(\gamma N)},
$$
in which $r_h$ is the half-mass radius, $N$ is the number of stars,
$m$ is the (mean) stellar mass, and $\gamma$ is a numerical constant
of order unity.  It follows that 
$$
N_{esc} = \frac{2\pi\mu}{0.138}\log(\gamma
N)\left(\frac{x_L}{r_h}\right)^{3/2}\left(\frac{4\Omega^2}{\kappa^2}-1\right)^{1/2}.
$$
The point of this result is not so much to furnish a numerical
estimate, but to point out that it is only weakly dependent on $N$.
In this sense a single numerical simulation with large $N$ is not more useful
for investigating the distribution along a tidal tail than a
single small simulation.  However, relaxation proceeds by the
cumulative effect of many small encounters, and the mean change (in
the energy of a star) per encounter decreases as $N$ increases.
Therefore the assumption of zero velocity at the moment when the star
crosses the tidal radius becomes increasingly accurate as $N$
increases, and we may expect that the density enhancements become
increasingly pronounced in this limit.

\section{Numerical verification}\label{sec:numerical}
\begin{figure*}
  \centering
  \includegraphics[width=120mm]{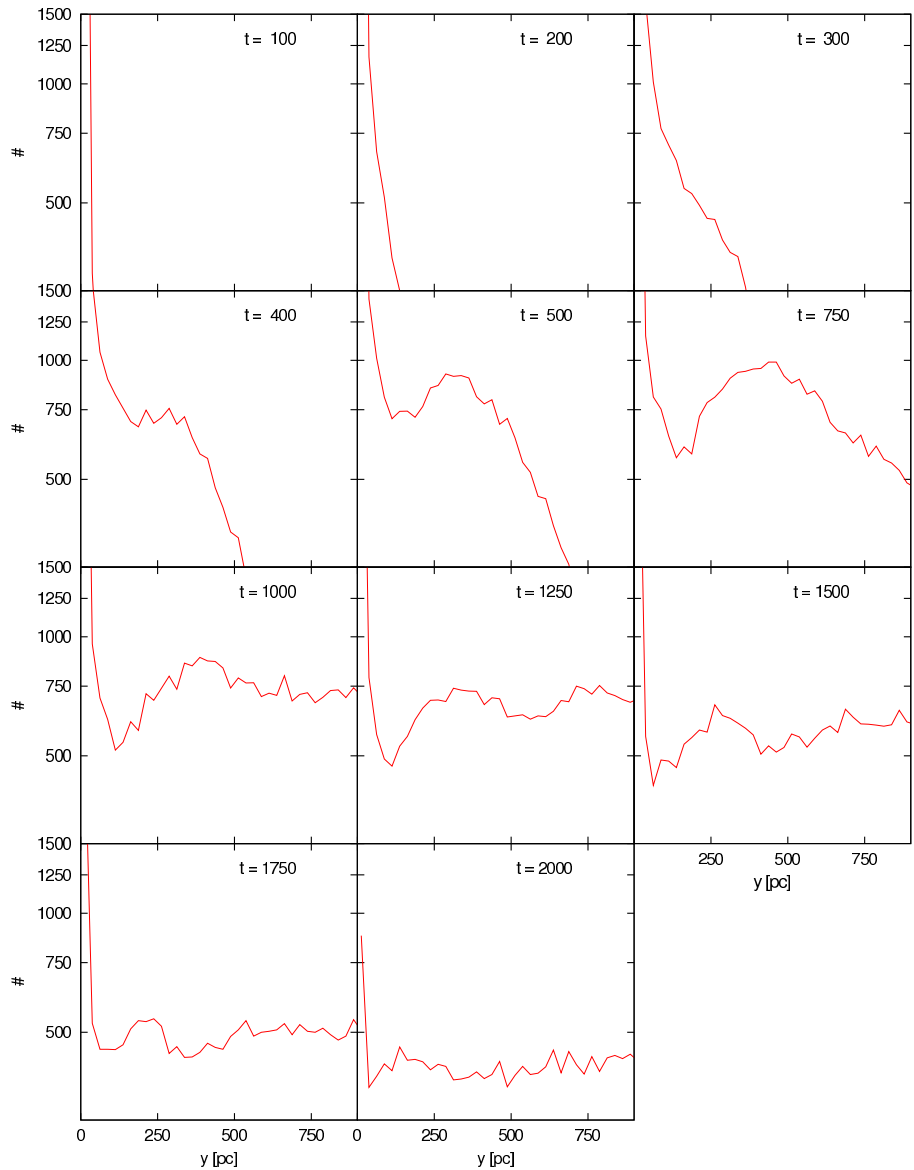}
  \caption{Time evolution of the number density of stars along the tidal
   tails. Since the leading and trailing tail are symmetric they are folded together. Time is given in Myr. As expected the overdensities need more
   than $t_C$ to build up; they are most pronounced at about
   $2-3\,t_C$. The predicted location of $y_C$ agrees fairly well with
   the results of the simulations. At $4\,t_C$ ($\simeq$1000 Myr) the
   cluster has lost about 70\% of its mass, and hence $y_C$ moves
   closer to the cluster centre.  Also the flux of escaping stars diminishes. At 2000 Myr the cluster has dissolved almost completely.}
  \label{fig:folded}
\end{figure*}
 
\begin{figure*}
  \centering
  \includegraphics[width=120mm]{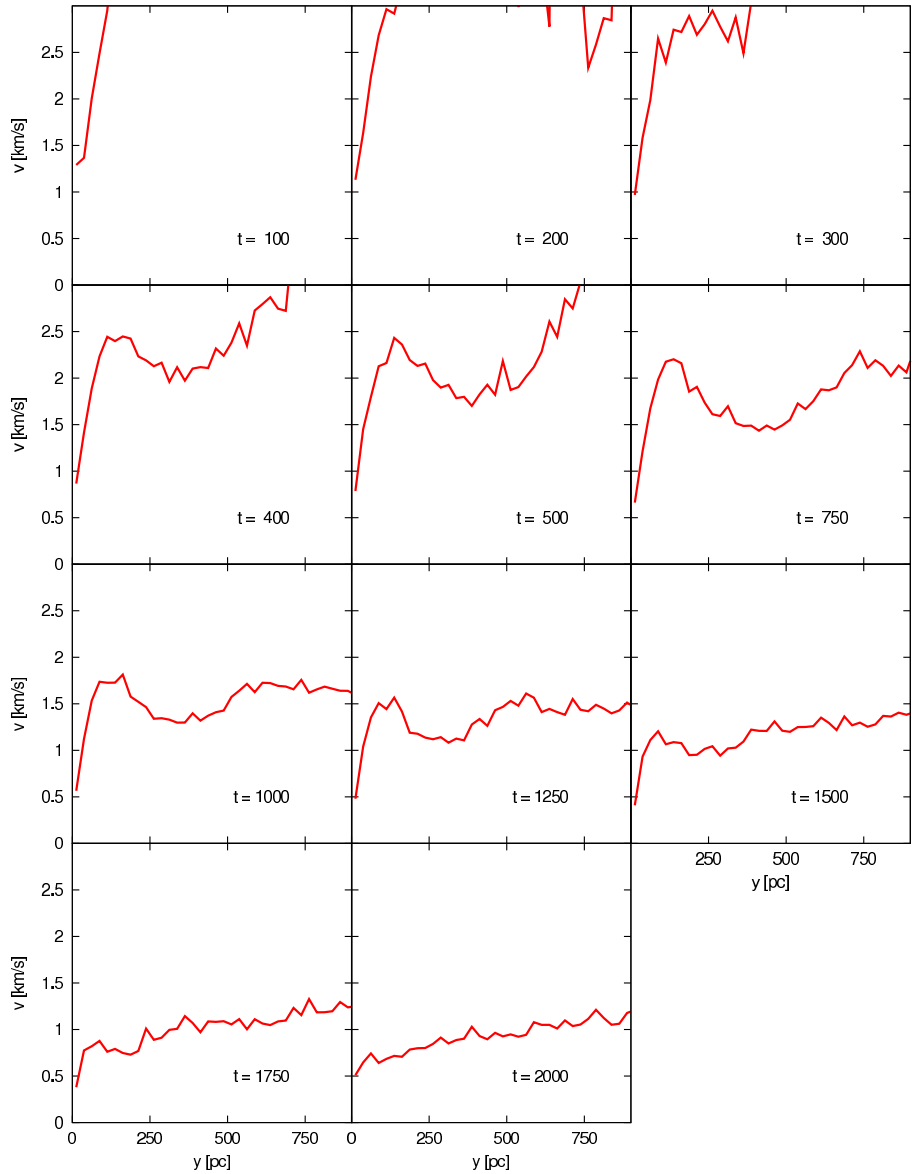}
  \caption{The same as Fig. \ref{fig:folded}, but here the mean speed of stars outside the
cluster (given in km/s) is shown.  In the beginning a few high-velocity
escapers dominate the outer parts.  As time goes by the speed profile
clearly shows the signature of the epicyclic motion described in Sec. \ref{sec:theory}
(i.e. the dip at about 300pc at t = 1250).  Furthermore the low speed
close to the cluster verifies the approximation that stars leave the
cluster with low velocities.  A similar plot is given in \citet{capuzzo2005}.}
  \label{fig:speed}
\end{figure*}

\begin{figure*}
  \centering
  \includegraphics[width=168mm]{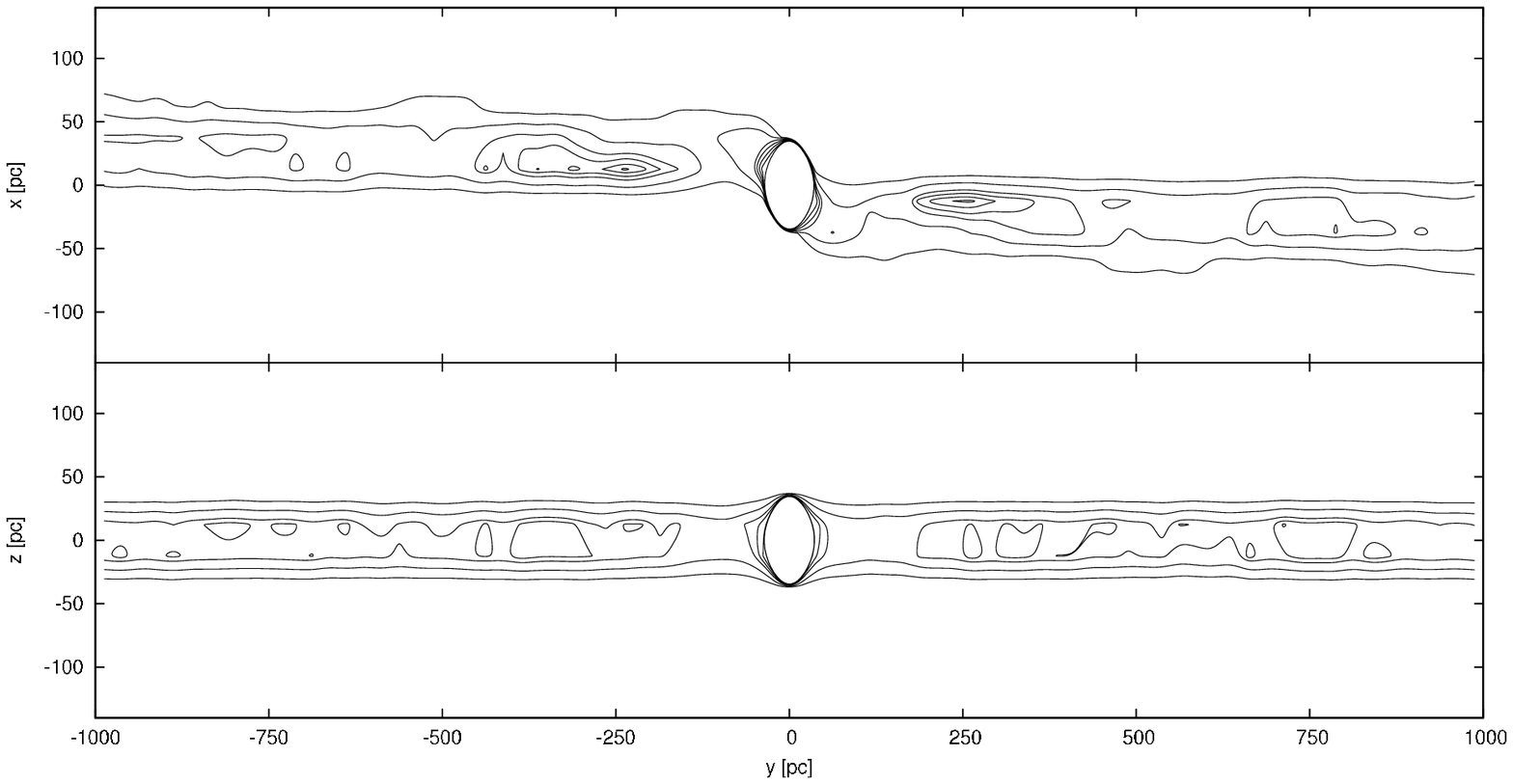}
  \caption{Contour plot of the surface density in the x-y (above) and z-y plane (below). The snapshot shows the superimposed test clusters at $t=1250$ Myr where the first and second peaks are visible best. Contours correspond to 50, 100, 150, 175, 200, 225 and 250 stars per bin, where the bin size was $25\,\mbox{pc} \times 25\,\mbox{pc}$.}
  \label{fig:contour}
\end{figure*}

The numerical simulations used to verify the existence of the predicted substructure were carried out with the collisional $N$-body code NBODY6 by Sverre Aarseth \citep{Aarseth99} on the 52-processor Sun Fire E15k at the Edinburgh Parallel Computing Centre. We used the supercomputer as a work farm to compute a set of 64 renditions of our test model which were then coadded to gain statistical accuracy.

The test model is a cluster of 1000 stars of each 1 $\mbox{M}_\odot$, following a Plummer density profile with a half-mass radius of 0.8 pc. The cluster orbits about a point mass galaxy of $M_G=9.565\cdot 10^{10} \mbox{M}_\odot$ at a galactic distance of $R_G=8.5$ kpc, which corresponds to a rotational velocity of $v_{rot}= 220$ kms$^{-1}$. 

In this particular case of a point-mass potential the tidal radius, $x_L$, is given by \citep{Spitzer87}
$$
x_L = R_G \left(\frac{M}{3M_G}\right)^{1/3},
$$
which yields an initial value of 12.9 pc for the above given
values. Therefore the cusps of the orbits of escaping stars are
initially located at $\pm y_C$, where
$$
y_C = 12 \pi x_L \simeq 490\,\mbox{pc},
$$
but this decreases with the ongoing loss of cluster mass. The epicyclic frequency of the cluster's orbit about the galactic centre is given by
$$
\kappa = \Omega = v_{rot}/R_G = 0.026\, \mbox{Myr}^{-1}.
$$
Hence the time the first stars need to reach this first cusp, $t_C$, is about 240 Myr. Before this time we do not expect any overdensity within the growing tidal tails.

The simulations were conducted until total dissolution, which occurred
at about 2000 Myr. For several times the positions of the
stars were first centred on the density centre of the remaining
cluster stars within the tidal radius. This was done for each
rendition before they were combined. Then the number density of stars
along the direction of the tidal tails, i.e. the y-axis, was counted in bins of 25
pc.

In Fig. \ref{fig:folded} the time evolution of the number density of
stars along the tidal tails is shown. Due to the symmetry of the leading and trailing tail they have been folded together to increase the statistical accuracy. As predicted in Section 2 the first stars need at least the time $t_C$ to reach the first cusp of
their oscillatory orbit along the tail at $y_C$. At $2\,t_C$ the
overdensity has built up completely and has become very pronounced. As
time goes by, the loss of mass causes a decrease in $x_L$, and by
eq.(\ref{eq:yc}) the location of the overdensity  moves closer to the
cluster.   At $4\,t_C$ ($\simeq 1000$ Myr) the cluster mass is only
about 300 $\mbox{M}_\odot$, hence $x_L$ has decreased to about 8.6 pc and so
$y_L$ is supposed to be at about 320 pc at $4\,t_C$ which agrees
fairly well with the numerical results.   Different simulations lose mass at different
rates, and the resultant distribution of values of $x_L$ (and hence
$y_C$) contributes to the smoothing out of the 
density maxima as $t$ increases.  Nevertheless, there are signs of
density extrema also at about $\pm2y_C$.

In Fig. \ref{fig:speed} the evolution of the mean speed outside the cluster is shown. The velocity change as described in Sec. \ref{sec:theory} is fairly well reproduced. In the beginning the mean speed is dominated by a few high-velocity escapers. But after building up, the overdensities correspond to a minimum in the mean speed while maxima correspond to underdensities. Here again the second peak is visible. Furthermore the initial rise of speed verifies the assumption that most stars leave the cluster with small velocities.

Fig. \ref{fig:contour} shows a snapshot of the surface density along the tidal tails after 1250 Myr. In both the x-y and y-z plane the over- and underdensities are fairly  pronounced, including (especially in the upper diagram) the additional overdensities at $\pm 2y_C$.

\section{Conclusions}

We have considered the longitudinal structure of the tidal tails
generated by stars escaping from a star cluster in a circular galactic
orbit.  Even though the tidal field is steady (in a rotating frame in
which the cluster centre is at rest), the tails exhibits density
enhancements. These correspond to places where escaping stars slow down in their epicyclic motion away from the star cluster.  Because stars escape with a somewhat limited
range of initial positions and velocities (when they cross the tidal
boundary) the location at which this happens is similar for most
escapers, and the resulting density enhancement is visible in
simulations provided that sufficient numbers of simulations are
coadded to improve the sampling statistics.  We have conducted such
simulations, using stars of equal mass.  The distance of the main
clump is around 38 tidal radii, in the case of a galaxy
modelled as a central point mass.

Similar clumps have been observed by \citet{capuzzo2005}, but those examples were a little ambiguous because (i) the galactic potential was triaxial, (ii) the orbit was not quite
circular, and (iii) the code was collisionless.  For these reasons the
mechanism of escape in their model was not clear, and could include an
unquantified episodic contribution from the time-dependence of the
tidal field.  Nevertheless they showed that the clumps coincide with
places where the escaping stars slow down. The main additional
contribution of the present paper is that we have identified the
mechanism causing those overdensities, by placing the problem in a more
idealised setting and providing a theoretical explanation.

Whether such density enhancements would be visible in the tidal tails
of real star clusters is complicated by their mass spectrum and,
especially for globular clusters, by the non-circularity of their
orbits. Nevertheless our results show that the density enhancements in tidal tails, which are apparently observed, should not necessarily be taken to indicate time-dependent tides, or encounters with perturbers such as giant molecular clouds, dark matter clumps or spiral arms.

\section*{Acknowledgements}
AM thanks the School of Mathematics at the University of Edinburgh for provision of a summer vacation scholarship, funded by the William Manson Prize,  when this research
was begun in 2006.

The simulations reported here were carried out by AHWK while a visitor to Edinburgh
under the HPC-EUROPA project (RII3-CT-2003-506079), with the support of the European Community - Research Infrastructure Action under the FP6 ``Structuring the European Research Area'' Programme.

The authors would like to thank the referee for fruitful suggestions.

\bsp

\label{lastpage}

\begin{thebibliography}{99}

\bibitem[\protect\citeauthoryear{Aarseth}{1999}]{Aarseth99} 
Aarseth S.~J., 1999, PASP, 111, 1333 

\bibitem[\protect\citeauthoryear{Belokurov et 
al.}{2006}]{belokurov2006} Belokurov V., Evans N.~W., Irwin
  M.~J., 
Hewett P.~C., Wilkinson M.~I., 2006, ApJ, 637, L29 

\bibitem[\protect\citeauthoryear{Capuzzo Dolcetta, Di Matteo, \& 
Miocchi}{2005}]{capuzzo2005} Capuzzo Dolcetta R., Di Matteo
  P., 
Miocchi P., 2005, AJ, 129, 1906 

\bibitem[\protect\citeauthoryear{Chandrasekhar}{1942}]{Chandrasekhar42} 
Chandrasekhar S., 1942, Principles of Stellar Dynamics (Chicago, Ill.: The University of Chicago Press)  

\bibitem[\protect\citeauthoryear{Choi, Weinberg, \& 
Katz}{2007}]{choi2007} Choi J.-H., Weinberg M.~D., Katz N., 
2007, MNRAS, 381, 987 

\bibitem[\protect\citeauthoryear{Chumak \& 
Rastorguev}{2006}]{chumak2006} Chumak Y.~O., Rastorguev
  A.~S., 
2006, AstL, 32, 157 

\bibitem[\protect\citeauthoryear{Combes, Leon, \& 
Meylan}{1999}]{combes1999} Combes F., Leon S., Meylan G.,
  1999, 
A\&A, 352, 149 

\bibitem[\protect\citeauthoryear{Dehnen et al.}{2004}]{Dehnen04} 
Dehnen W., Odenkirchen M., Grebel E.~K., Rix H.-W., 2004, AJ, 127, 2753

\bibitem[\protect\citeauthoryear{di Matteo, Capuzzo Dolcetta, \& 
Miocchi}{2005}]{dimatteo2005} di Matteo P., Capuzzo Dolcetta
  R., 
Miocchi P., 2005, CeMDA, 91, 59 

\bibitem[\protect\citeauthoryear{Fellhauer et 
al.}{2007}]{fellhauer2007} Fellhauer M., Evans N.~W., Belokurov
  V., 
Wilkinson M.~I., Gilmore G., 2007, MNRAS, 380, 749 

\bibitem[\protect\citeauthoryear{Grillmair et 
al.}{1995}]{grillmair1995} Grillmair C.~J., Freeman K.~C., Irwin
  M., 
Quinn P.~J., 1995, AJ, 109, 2553 

\bibitem[\protect\citeauthoryear{Hayli}{1970}]{hayli1970} Hayli 
A., 1970, A\&A, 7, 17 

\bibitem[\protect\citeauthoryear{Johnston, Sigurdsson, \& Hernquist}{1999}]{johnston1999} Johnston K.~V., Sigurdsson S., Hernquist L., 1999, MNRAS, 302, 771

\bibitem[\protect\citeauthoryear{Kharchenko, Scholz, \& 
Lehmann}{1997}]{kharchenko1997} Kharchenko N., Scholz R.-D.,
  Lehmann 
I., 1997, A\&AS, 121, 439 

\bibitem[\protect\citeauthoryear{Koch et
    al.}{2004}]{koch2004} 
Koch A., Grebel E.~K., Odenkirchen M., Mart{\'{\i}}nez-Delgado D.,
    Caldwell 
J.~A.~R., 2004, AJ, 128, 2274 

\bibitem[\protect\citeauthoryear{Lee et
    al.}{2003}]{lee2003} 
Lee K.~H., Lee H.~M., Fahlman G.~G., Lee M.~G., 2003, AJ, 126, 815 

\bibitem[\protect\citeauthoryear{Lee et
    al.}{2004}]{lee2004} 
Lee K.~H., Lee H.~M., Fahlman G.~G., Sung H., 2004, AJ, 128, 2838 

\bibitem[\protect\citeauthoryear{Lee, Lee, \& 
Sung}{2006}]{lee2006} Lee K.~H., Lee H.~M., Sung H., 2006, 
MNRAS, 367, 646 

\bibitem[\protect\citeauthoryear{Lehmann \& 
Scholz}{1997}]{lehmann1997} Lehmann I., Scholz R.-D., 1997,
  A\&A, 
320, 776 

\bibitem[\protect\citeauthoryear{Leon, Meylan, \& 
Combes}{2000}]{leon2000} Leon S., Meylan G., Combes F.,
  2000, 
A\&A, 359, 907 

\bibitem[\protect\citeauthoryear{Montuori et 
al.}{2007}]{montuori2007} Montuori M., Capuzzo-Dolcetta R., Di 
Matteo P., Lepinette A., Miocchi P., 2007, ApJ, 659, 1212 

\bibitem[\protect\citeauthoryear{Odenkirchen et 
al.}{2001}]{Odenkirchen01} Odenkirchen M., et al., 2001, ApJ, 548, 
L165 

\bibitem[\protect\citeauthoryear{Odenkirchen et 
al.}{2003}]{odenkirchen2003} Odenkirchen M., et al., 2003, AJ,
  126, 
2385 

\bibitem[\protect\citeauthoryear{Rockosi et 
al.}{2002}]{rockosi2002} Rockosi C.~M., et al., 2002, AJ, 124,
  349 

\bibitem[\protect\citeauthoryear{Sohn et
    al.}{2003}]{sohn2003} 
Sohn Y.-J., et al., 2003, AJ, 126, 803 

\bibitem[\protect\citeauthoryear{Spitzer}{1987}]{Spitzer87} 
Spitzer L., 1987, Dynamical Evolution of Globular Clusters (Princeton: Princeton University Press)

\bibitem[\protect\citeauthoryear{Yim \&
    Lee}{2002}]{yim2002} 
Yim K.-J., Lee H.~M., 2002, JKAS, 35, 75 

\end{thebibliography}
\end{document}